# Signal Processing Techniques to Reduce the Limit of Detection for Thin Film Biosensors

†Simon J. Ward*, ‡Rabeb Layouni, §Sofia Arshavsky-Graham, §Ester Segal, and †Sharon M. Weiss.

†Department of Electrical Engineering and Computer Science, Vanderbilt University, Nashville, Tennessee 37235, USA.
‡Department of Chemical and Biomolecular Engineering, Vanderbilt University, Nashville, Tennessee 37235, USA
§Department of Biotechnology and Food Engineering, Technion Israel Institute of Technology, Technion City, 32000 Haifa, Israel.

**ABSTRACT:** The ultimate detection limit of optical biosensors is often limited by various noise sources, including those introduced by the optical measurement setup. While sophisticated modifications to instrumentation may reduce noise, a simpler approach that can benefit all sensor platforms is the application of signal processing to minimize the deleterious effects of noise. In this work, we show that applying complex Morlet wavelet convolution to Fabry-Pérot interference fringes characteristic of thin film reflectometric biosensors effectively filters out white noise and low frequency reflectance variations. Subsequent calculation of an average difference in phase between the filtered analyte and reference signals enables a significant reduction in the limit of detection (LOD) enabling closer competition with current state-of-the-art techniques. This method is applied on experimental data sets of thin film porous silicon sensors (PSi) in buffered solution and complex media obtained from two different laboratories. The demonstrated improvement in LOD achieved using wavelet convolution and average phase difference paves the way for PSi optical biosensors to operate with clinically relevant detection limits for medical diagnostics, environmental monitoring, and food safety.

Detecting low levels of an analyte is crucial for a wide variety of sensing applications, including medical diagnostics in which early detection of low concentrations of disease biomarkers enables preventative measures and treatments with exponentially higher rates of positive patient outcomes[1], and environmental monitoring in which detection of toxic substances harmful to humans at ultra-low levels can save lives[2]. Of the many ways to improve limit of detection (LOD), arguably the cheapest and easiest strategy is signal processing, which can improve detection limits by removing various noise signatures. Signal processing for noise reduction is applicable to all systems, from costly, high resolution and highly accurate laboratory instruments to compact, simple and cheap point-of-care devices.

Signal processing has been used in a wide variety of biosensing platforms to enhance sensitivity, achieve lower detection limits, and provide deeper insight into the physical properties of the sensor and the sensing mechanism. For surface plasmon resonance sensors, one of the most successful optical biosensing platforms, techniques such as polynomial fitting[3], centroid detection[4] and tracking[5], complemented by dynamic baseline algorithms[6], linear data analysis[7], locally weighted parametric regression[8], and the radon transform technique[9] have been used for higher resolution resonance detection and reduced noise susceptibility. The performance of quartz crystal microbalance sensors, another effective and widely used biosensing platform, has also benefitted from signal processing methods, such as the fractional Fourier transform[10], heterodyning[11], and moving average and Savitzky-Golay filtering[12]. Additionally, frequency locking along with nonlinear filtering has been used to lower the noise floor of microtoroid optical biosensors, enabling single molecule detection[13].

In this work, we introduce a signal processing approach based on Morlet wavelet convolution and phase analysis that can dramatically lower the detection limit of thin film biosensors, which are among the simplest optical biosensor platforms. We utilize porous silicon (PSi) thin film biosensors for our demonstration of the benefits of this signal processing approach. Because the characteristic pore size of PSi films is much smaller than the wavelength of light, PSi is treated as an effective medium, and the optical thickness and refractive index are referred to as the effective optical thickness and effective refractive index, respectively[14]. PSi biosensors have an extraordinarily large capacity for biomolecule adsorption due to their vast surface area and strong interaction between light and adsorbed molecules. However, the lowest LODs achieved in the literature using PSi biosensors typically are higher than those based on other biosensing systems[15,16], in part due to the contributions of noise signatures which have not been mitigated, as well as mass transport challenges that have been recently investigated in other work[17,18].

The reflectance spectrum of a thin film contains a series of approximately sinusoidal Fabry-Pérot interference fringes. One common signal processing approach to measure changes in these fringes due to adsorbed biomolecules is the reflective interferometric Fourier transform spectroscopy (RIFTS) method[19], which involves determining the dominant frequency of these fringes by carrying out a Fast Fourier Transform (FFT) on the reflectance spectra, and then identifying the peak interference fringe frequency. This method gives important physical information about the optical properties of the film, but its accuracy is susceptible to noise. More recently, promising improvements in LOD of PSi thin films have been reported[20,21] using the Interferogram Average over Wavelength (IAW) technique. However, the robustness of the IAW method is an issue when there are offset and amplitude variations in experimentally measured data. These noise sources, found in most data sets, are caused by the measurement system as well as scattering of light by molecules adsorbed to the surfaces of the PSi thin film and to the surfaces of microfluidic channels when they are utilized[22]. Moreover, noise contributions tend to be more pronounced in sensing assays carried out in complex media.

Since the Fabry-Pérot fringes characteristic of thin film reflectance spectra are approximately sinusoidal, they are well suited to band pass filtering designed to remove undesired low frequency variations as well as white noise at all frequencies above and below the pass band, which improves robustness and increases sensitivity, respectively. In this work, we show that application of complex Morlet wavelet convolution, which has previously been used to filter electrical signals generated from electrodes monitoring brain or heart activity[23] and seismic activity[24], is a highly effective approach to reduce noise signatures in optical spectra of thin films such as PSi single layer biosensors. We coin the name, Linear Average Morlet Phase (LAMP) method, to describe our approach that utilizes both a complex wavelet and extracted phase information. The LAMP technique enables a reduction of the LOD by almost an order of magnitude relative to RIFTS and IAW, and makes thin film sensors, such as those based on PSi, potentially viable options for many clinical applications. We note that the LAMP signal processing technique does not negatively impact manufacturability, cost, complexity, or response time, and can be applied alongside other sensitivity enhancing techniques.

## Signal Processing Approaches

Several different signal processing approaches can be applied to help interpret measured optical spectra from thin films. For many applications, including biosensing, there is particular interest in being able to determine small changes between two measured thin film optical spectra. The reflectance spectrum of a thin film is characterized by Fabry-Pérot fringes (e.g., Figure 1a(i)), which are caused by interference from successive reflections from the front and back faces of the film. The reflectance of a prototypical Fabry-Pérot etalon, $R_{tf}$, is given in Eqn. (1).

$$R_{tf} = \frac{2R(1 - cos\phi)}{(1 - 2Rcos\phi + R^2)} \quad (1)$$

In this equation, $R$ is the square of the reflection coefficient at a single interface (i.e., $R = |r|^2$ where $r$ is the Fresnel amplitude reflection coefficient). The round-trip phase shift is given by $\phi = \frac{4\pi nL}{\lambda}$, where $n$ is the refractive index of the thin film, $L$ is the thickness of the film which must be less than the coherence length of the incident light, and $\lambda$ is the wavelength of light. The derivation of Eqn. (1) assumes negligible absorption, and that the front and back sides of the thin film interface with the same material. The latter assumption is not strictly valid for many thin film sensors, which often interface with a solution of biomolecules at the front surface and a substrate of different refractive index at the back surface. Accordingly, for many thin film sensors, the intensity of reflected light will be modified from the ideal Fabry-Pérot etalon case and there may be a $\pi$ phase shift in the Fabry-Pérot fringes. Nevertheless, the relationship indicated by Eqn. (1) that minima in reflectance spectra occur at $\phi = 2\pi m$, where $m = 0, 1, 2 ...$, holds for all thin film sensors. This phase relationship corresponds to wavelengths that satisfy the relation, $2nL = m\lambda$. Hence, plotting $R_{tf}$ against inverse wavelength (i.e., wavenumber), will yield a series of equally spaced sinusoidal fringes if no external noise sources are considered. The frequency of these fringes is 2nL, which is known as the optical thickness. We note that since the independent variable – wavenumber – is in units of inverse distance, the frequency of the reflectance fringes is given, in this case, in units of distance.

### RIFTs and IAW Methods

One approach to characterizing the reflectance spectra of thin films is to measure optical thickness by monitoring the frequency of the Fabry-Pérot fringes in wavenumber-space. We note that it is essential to plot the reflectance of the film as a function of wavenumber so that the Fabry-Pérot fringes are equally spaced across the spectrum. Changes in optical thickness of thin films are characterized by changes in the frequency of these Fabry-Pérot fringes. The RIFTS method aims to identify the dominant frequency of the Fabry-Pérot fringes, corresponding to the optical thickness, from the FFT of the reflectance spectra. The RIFTS technique consists of the following four operations:

1. Plot the reflectance spectra against wavenumber using cubic spline interpolation to generate equally spaced points (Figure 1 (a) (i)).
2. Apply a Hann window to the chosen spectral measurement range to enforce periodicity and consequently reduce spectral leakage (Figure 1 (a) (ii)).
3. Increase the length of the data by zero padding to realize the desired resolution of 2nL values when the FFT is applied (Figure 1 (a) (ii)). Note that the highest efficiency is achieved when the number of points is a power of two.
4. Carry out an FFT and identify the frequency of the dominant peak (Figure 1 (a) (iii)).

Changes in optical thickness caused by biomolecule attachment can therefore be determined in a straightforward manner, by monitoring the change in the peak frequency result from the RIFTs method applied to spectra measured before and after exposure of the thin film to biomolecules.

Since a change in Fabry-Pérot fringe frequency causes the Fabry Pérot fringes to shift (by an amount proportional to wavenumber), another strategy to measure biomolecule attachment is to calculate the difference between spectra before and after biomolecule exposure. The IAW method uses this approach, and is implemented as follows:

1. Find the difference between a 'reference' reflectance spectrum before exposure of the target biomolecule, and an 'analyte' spectrum after biomolecule exposure, termed the 'interferogram' (Figure 1 (b) (i)).
2. Zero the interferogram by subtracting the mean from each value (Figure 1 (b) (ii)).

3. Integrate the absolute value of each point in the interferogram over the chosen spectral measurement range (Figure 1 (b) (iii)).

It should be noted that the magnitude of the IAW signal change is strongly dependent on the changing amplitude of the Fabry Pérot fringes at each given wavenumber in the reference and analyte spectra, which in turn makes the technique more susceptible to noise contributions that affect the measured amplitude of the spectra. The RIFTS technique, on the other hand, is not strongly affected by the relative amplitudes of the spectra.

### Complex Morlet Wavelet Convolution: LAMP method

Because the lowest achievable LOD for thin film sensors is limited by both high frequency white noise and low frequency noise signatures in the reflectance spectrum, band pass filtering is a promising signal processing approach to apply. In particular, complex Morlet wavelet convolution is well suited for filtering the spectra of thin film sensors as a means of maximally removing noise while retaining the desired optical signal. A complex Morlet wavelet is a complex exponential with a Gaussian envelope, which results in a Gaussian line shape in frequency space. The convolution of a Morlet wavelet with a signal is equivalent to multiplication in frequency space, which can act as a strict band pass filter. Since the wavelet is complex, the filtered result is a complex signal from which phase and amplitude can be extracted. The first stage of the LAMP method, complex Morlet wavelet band pass filtering, is summarized by the following steps:

1. Plot reflectance spectra against wavenumber, using linear interpolation so points are equally spaced (Figure 1 (c) (i)).
2. Zero pad the data to obtain the desired resolution in frequency space.
3. Carry out an FFT and identify the center frequency and full width at half maximum (FWHM) of the dominant peak (Figure 1 (c) (ii)).
4. Use the center frequency and FWHM of the dominant peak to define a complex Morlet wavelet, shown in the green trace in Figure 1 (c) (i) in wavenumber space. An FFT of the complex Morlet wavelet is shown in the green trace in Figure 1 (c) (ii) and reveals the band pass filtering ability of the Morlet wavelet. The spacing between wavenumber values of the wavelet should be the same as for the interpolated reflectance spectra.
5. Convolve the wavelet with the reflectance data (Figure 1 (c) (i)) to obtain the filtered output shown in Figure 1 (c) (iii).
6. Extract the phase from the real and imaginary components of the filtered output over the spectral measurement range, as shown in Figure 1 (c) (iii) alongside the real part of the filtered output. The amplitude can also be extracted from the complex output from the filtering stage and can be used to normalize the Fabry-Pérot fringes (Figure 1 (c) (iii), (iv)).

By carrying out a FFT of the normalized fringes, it becomes clear in frequency space that the high and low frequency noise components have been removed from the filtered data (Figure 1 (c) (v)). For the LAMP method, only the phase is used, as further processing was found to introduce additional noise; however, for some applications, it may be useful to obtain normalized fringes.

The first three steps of complex Morlet wavelet band pass filtering are calculated in a similar way to the RIFTS method, with the exception that no Hann window is applied to avoid broadening the dominant frequency peak (i.e., to give a more accurate FWHM of the frequency peak that is used in defining the wavelet), at the expense of having greater spectral leakage and extra lobes introduced.

The LAMP method uses complex Morlet wavelet filtering to measure changes in optical thickness as follows:

1. Apply complex Morlet wavelet band pass filtering, as described previously, to a 'reference' reflectance spectrum before exposure of the target biomolecules and to an 'analyte' spectrum after biomolecule exposure (Figure 1 (d) (i)). The real part of the filtered output is shown in Figure 1 (d) (ii).
2. Extract the phase from the complex result of filtering, and unwrap it by adding $2\pi$ every time the phase goes through another complete cycle to give a linearly increasing phase for both the reference and analyte spectra (Figure 1 (d) (ii)).
3. Correct for phase of reference and analyte spectra starting in different cycles. For example, if the initial phase of the reference spectra at the lowest wavenumber in the measurement window has a phase just below 2pi, biomolecule adsorption could cause the initial phase to shift into the next cycle with a phase just above 0 for the analyte spectra, which causes an additional 2pi difference between the reference and analyte phase at every value of wavenumber, unrelated to biomolecule adsorption. To correct for this difference in the initial phase of the reference and analyte spectra, a coarse value of initial phase is estimated using the FFT dominant fringe frequency, alongside the accurate Morlet Wavelet acquired phase, to determine when the phase has entered a different cycle.
4. Average the difference between the unwrapped linear phase for the reference and target spectra, which yields the LAMP result (Figure 1 (d) (iii)).

To optimize the performance of all three signal processing methods (RIFTS, IAW, and LAMP), the spectral measurement range to analyze should be chosen to include as much of the measured spectra as possible while rejecting the extremes of the spectra with poor signal to noise ratio (S/N). Hence, the most appropriate spectral range for the signal processing methods is dependent on the limitations of the optical components and spectrometer utilized in the measurements. In this work, a spectral measurement range of 500-800 nm was used unless otherwise noted in order to facilitate comparison between techniques and previous studies.

### Computational Generation of Spectra

To provide a comprehensive comparison between different signal processing techniques, reflectance spectra of a single layer PSi thin film were simulated using the transfer matrix method. White Gaussian noise was added to the simulated spectra to better mimic measured data. The simulated thin film was defined as a 2.4 µm thick layer with effective refractive index of 1.2, which are realistic parameters comparable with experimental data sets. To rigorously test each signal processing method and mirror noise encountered in real experimental spectra as closely as possible while keeping the analysis simple and generalizable, three different noise contributions were considered: (1) a simple shift of the Fabry-Pérot fringes due to a refractive index change with added white Gaussian noise, (2) a refractive index change with white Gaussian noise and a unitary wavelength dependent linear offset applied to the Fabry-Pérot fringes, and (3) a refractive index change with white Gaussian noise and a unitary wavelength-dependent linear amplitude variation applied to the Fabry-Pérot fringes. The S/N when comparing the signal to the white Gaussian noise was 27.7 dB. The S/N decreased to 7.9 dB and 7.7 dB when the linear offset was superimposed or the linear amplitude variation was introduced, respectively. A linearly varying offset and amplitude are simple modifications to fringes with a relatively complex spectral composition, testing the signal processing methods immunity to a wide variety of noise signatures seen experimentally.

### Limit of Detection

The LOD of a biosensor is the minimum concentration or number of molecules that can be reliably detected. A normal distribution of measurements can be built up both before and after biomolecule exposure, termed the blank and shifted distributions respectively. The LOD is typically quantified as the concentration of target molecule leading to a response such that the lower bound of the highest 5% of the blank distribution, which lies 1.65 standard deviations above the mean, coincides with the upper bound of the lowest 5% of the shifted distribution, which lies 1.65 below the mean[25]. Equivalently stated,

$$\mu_{shift} - \mu_{blank} = 1.65\sigma_{blank} + 1.65\sigma_{shift} \quad (2)$$
$$\approx 3.3\sigma_{blank}$$

where $\mu_{blank}$ and $\mu_{shift}$ are the mean of the blank and shifted distributions, respectively, and $\sigma_{blank}$ and $\sigma_{shift}$ are the standard deviations of the blank and shifted distributions, respectively. The standard deviations are typically assumed to be approximately equal. When introducing the offset or amplitude gradient, the resulting change in response was isolated by calculating the blank distribution with and without the gradient at a constant refractive index. The effect on the response of each signal processing technique when applying a gradient to the offset or amplitude, $\Delta g$, was added to the blank-distribution standard deviation $\sigma_{blank}$, resulting in a LOD of $3.3(\sigma_{blank} + \Delta g)$. Including this effect as part of the standard deviation of the blank distribution enables modeling a distribution of offset or amplitude gradients that occur in tandem with any refractive index change and white noise-induced interference fringe shifts.

The mean and standard deviation of the blank and shifted distributions, as well as the blank distributions incorporating an offset or amplitude gradient, were each calculated from 1000 separately generated spectra with a random white noise contribution.

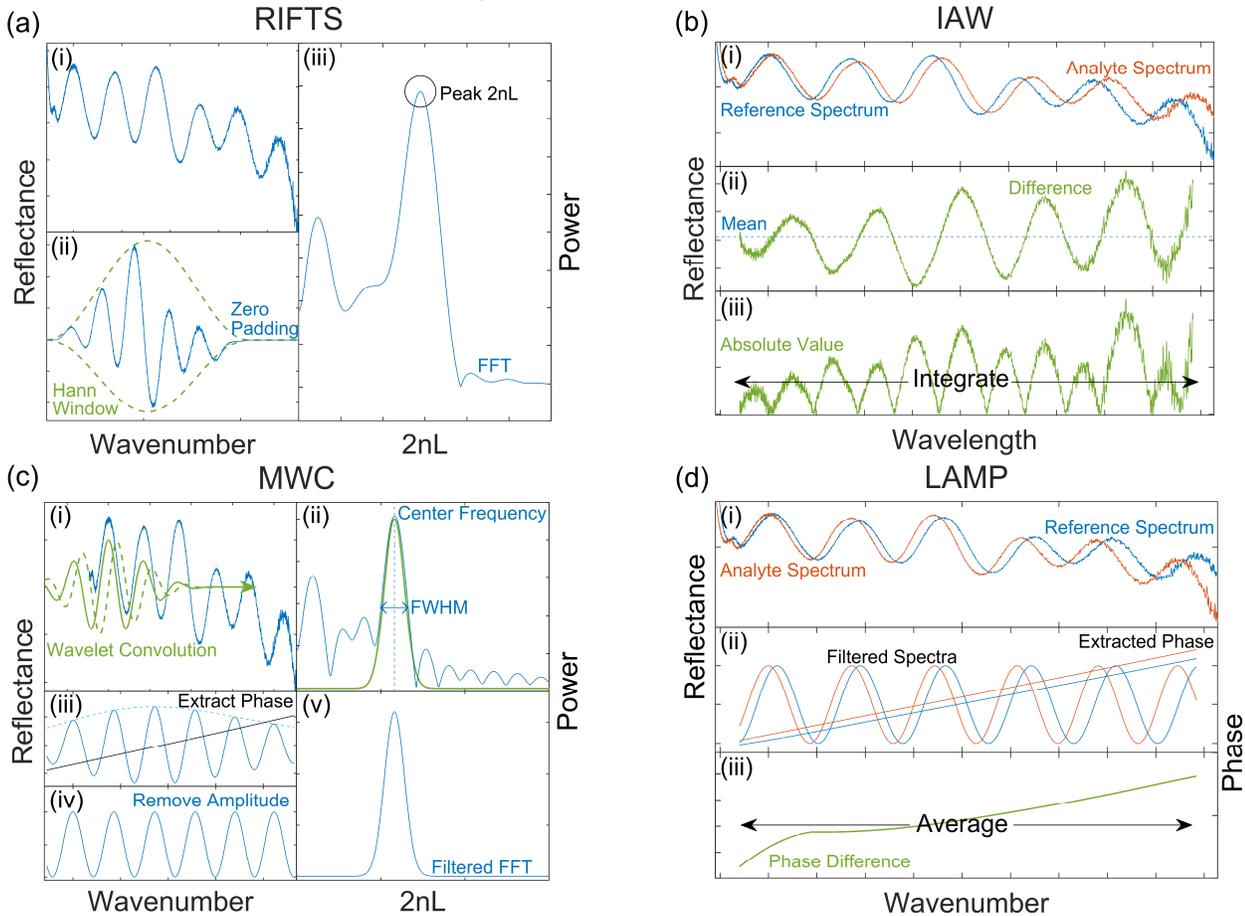

Figure 1. Illustration of signal processing techniques: (a) RIFTS, (b) IAW, (c) complex Morlet wavelet convolution, and (d) LAMP, applied to experimentally measured reflectance spectra of a single layer PSi thin film. The steps for each method are described in detail in the text.

## Experimental Methods

Signal processing approaches were applied to the measured data from two experimental systems: (1) non-specific bovine serum albumin (BSA) assay in buffer to investigate the LOD, and (2) anterior gradient 2 (AGR2) protein biosensor to investigate (i) the robustness of the signal processing approaches in buffered and complex media and (ii) selectivity in comparing target and non-target proteins. Non-specific adsorption of BSA is commonly utilized to demonstrate the proof-of-concept performance of biosensing platforms, and AGR2 is a biomarker for many types of cancer. These assays were executed in different labs (Vanderbilt University and Technion Israel Institute of Technology), with different experimental setups, different experimental procedures, and different materials, to demonstrate the broad applicability of the LAMP technique when encountering different noise fingerprints.

### BSA Assay (carried out at Vanderbilt)

*Materials*

Chemicals were all analytical grade, used without further purification. De-ionized (DI) water with resistivity 15 MΩ cm, from a Millipore Elix water purification system was used for all solutions. Single side polished, boron doped silicon wafers (⟨100⟩, 0.01–0.02 Ω cm, 500–550 μm) were purchased from Pure Wafer. Ethanol and BSA were purchased from Thermo Fisher Scientific, and pH 4 reference standard buffer, used for BSA solutions, was obtained from Sigma-Aldrich. Aqueous HF (48-51%) was purchased from Acros Organics. Solution pH was measured using a Mettler Toledo Seven Easy pH-meter.

*Fabrication of Single Layer PSi*

Single layer PSi thin films were fabricated by anodic etching of p-type silicon wafers with HF, described in detail elsewhere[14]. A solution of 15% HF in ethanol was used, along with an AMMT MPSB porous silicon wafer etching system. First, a sacrificial layer was etched with a current density of 70 mA cm$^{-2}$ for 100 s and then dissolved in 1 M KOH solution. Next, the sample was thoroughly cleaned with DI water and ethanol and then etched again at 70 mA cm$^{-2}$ for a further 50 s. Finally, the wafer was oxidized at 800°C for 10 min using a Lindberg/Blue M 1100°C box furnace to passivate the silicon by forming an insulating layer of $SiO_2$. The PSi films were 1.77 μm thick, with a porosity and average pore diameter of 67% and 49 nm, respectively, as determined by scanning electron microscopy (SEM) imaging.

*Optical Reflectance Measurements*

Light from a quartz tungsten halogen light source from Newport was coupled into an optical fiber, which is split before entering a lens tube housing and Olympus SPlan 10x microscope objective lens. The height of the lens relative to the PSi sample was adjusted to focus the light to a spot size of approximately 5 mm diameter on the surface of the PSi films. The reflected light was then collected by the same lens and coupled by another optical fiber into an Ocean Optics USB4000 CCD spectrometer, outputting the spectra to a PC running the Ocean Optics Spectra Suite software. 100 spectra were averaged and recorded once per second.

*Experimental Procedure*

PSi samples 5 mm × 5 mm were cleaved from the etched wafer, washed with water and ethanol, and dried under nitrogen. Reflectance spectra were then measured to obtain a baseline effective optical thickness and an IAW and LAMP reference spectrum. A further 100 measurements were taken to quantify the noise floor for each method: the blank distribution standard deviation $\sigma_{blank}$. Different concentrations of BSA were prepared in 80% pH 4 reference standard buffer, 20% DI water solutions (3 pM, 300 pM, 30 nM, 300 nM, 3 μM, 30 μM, 300 μM) and were drop cast on the samples and left to incubate for 2 hours. The samples were subsequently washed in a water bath for 10 s, which removes unbound and potentially a small number of weakly bound molecules in the pores and from the surface. The reflectance change after this wash step showed almost no dependence on wash duration. After washing, the PSi was dried under nitrogen and measured again. There were 16 repeats performed at each concentration. A number of adsorption isotherms were fit to the data using a least squares fit weighted by the variance due to the heteroscedasticity of the data, and the chi-squared goodness of fit was calculated to quantify quality of model fit and determine the most appropriate adsorption model. The predicted LOD was then calculated, given by the concentration where the best fit line intersects $3.3\sigma_{blank} + I$, where $I$ is the intercept of the line of best fit.

### AGR2 Biosensor (carried out at Technion)

*Materials*

Chemicals were all analytical grade, used without further purification. Double distilled water (ddH2O) with resistivity 18.2 MΩ·cm from a Milli-Q water purification system was used for all solutions. Heavily p-doped silicon wafers (⟨100⟩, 0.90–0.95 mΩ cm) were purchased from Sil'tronix Silicon Technologies. Aqueous 48% hydrofluoric acid (HF), and ethanol were purchased from Bio-Lab ltd. (3-Aminopropyl)triethoxysilane (APTES), diisopropylethylamine (DIEA), succinic anhydride, N-(3-Dimethylaminopropyl)-N′-ethylcarbodiimide hydrochloride (EDC), N-Hydroxysuccinimide (NHS), acetonitrile and all buffer salts were supplied by Merck. Anti-AGR2 aptamer sequence[26], 5'-TCT-CGG-ACG-CGT-GTG-GTC-GGG-TGG-GAG-TTG-TGG-GGG-GGG-GTG-GGA-GGG-TT-3', was purchased with a 5'-amino modification from Integrated DNA Technologies. AGR2 protein was purchased from MyBioSource Inc. Rabbit Immunoglobulin G (IgG) was purchased from Jackson ImmunoResearch Labs Inc. Human blood plasma from healthy subjects was purchased from Merck. Selection buffer (SB) was composed of 137 mM NaCl, 20 mM KCl, 10 mM Na2HPO4 and 2mM KH2PO4 (pH 7.4).

*Fabrication of Single Layer PSi*

Single layer PSi thin films were fabricated by anodic etching of p-type silicon wafers with HF. First, a sacrificial layer was etched with a current density of 300 mA cm$^{-2}$ for 30 s, using 3:1 (v/v) aqueous HF to ethanol solution, and then dissolved in 0.1 M NaOH solution, to remove surface impurities and oxides. Another layer is formed using identical conditions to the sacrificial layer, and the resulting PSi films were thermally oxidized at 800°C for 1 hour in a Thermo Scientific, Lindberg/Blue M™ 1200ºC Split-Hinge tube furnace. The PSi film thickness, average porosity and range of pore sizes were measured with a Carl Zeiss Ultra Plus high-resolution SEM at an accelerating voltage of 1 keV, and were found to be ~4.9 μm, 74%, and 40-60 nm, respectively.

*Surface Chemistry Preparation*

Amino-terminated aptamers, consisting of the anti-AGR2 sequence, were attached to the surface of the PSi films by carbodiimide coupling chemistry and amino silanization, as detailed elsewhere[27,28]. Briefly, the oxidized PSi films were amino-silanized in a solution of 1% APTES and 1% DIEA in ddH2O (v/v) for 1 h and annealed at 100°C for 15 min. Subsequently, surface carboxylation was carried out in a solution of 10 mg mL$^{-1}$ succinic anhydride and 2% (v/v) DIEA in acetonitrile for 3 h, followed by surface activation with 10 mg mL$^{-1}$ EDC and 5 mg mL$^{-1}$ NHS in 0.5 M MES buffer (pH 6.1) for 1 h. The activated surface was then reacted with the 50 µM amino-terminated aptamers for 1 h, followed by blocking with Tris buffer (50 mM, pH 7.4).

*Optical Reflectance Measurements*

For the biosensing experiments, the anti-AGR2 aptasensor was mounted in a custom-made Plexiglas cell. Light from a tungsten light source was coupled using a bifurcated optical fiber into an objective lens and focused onto the center of the PSi sample with a spot size of approximately 1 mm2. The reflected light was collected through the same fiber into an Ocean Optics USB4000 CCD spectrometer.

*Experimental Procedure*

Firstly, a baseline was established by incubating the aptasensor in SB. Next, an ARG2 protein solution either in SB or in 50% blood plasma in SB was incubated on top of the sample for one hour. Finally, the sample was washed with SB until equilibrium was reached and all unbound molecules were removed.

## Results and Discussion

### Simulation Results Comparing LOD of Different Signal Processing Techniques

The minimum detectable refractive index change of a thin film was calculated by applying different signal processing techniques to reflectance spectra simulated using the transfer matrix method with additive white Gaussian noise. In order to systematically evaluate the robustness of the signal processing techniques to noise contributions often arising during experimental measurements, the LOD analysis was carried out with (1) no added systematic noise contributions, (2) a linearly varying signal offset, and (3) a linearly varying amplitude increase. As shown in Table 1, the LAMP technique achieves a substantially lower LOD than both RIFTS and IAW methods in all cases.

We note that all the signal processing techniques perform proportionally better when the white noise is reduced. In order to determine this effect of S/N on the performance of the different signal processing techniques, the level of white Gaussian noise was varied by two orders of magnitude. A linear relationship was found between S/N and LOD for each signal processing technique, such that the relative merit of the techniques shown in Table 1 can be considered independent of Gaussian noise level. The RIFTS method performs relatively poorly in the presence of offset variations since the maximum of the dominant fringe frequency peak is shifted by the superposition of lower frequency components. On the other hand, the LOD for RIFTS is almost unchanged for amplitude variations, which only change the power of the dominant fringe frequency peak rather than its position. The IAW method is negatively affected by both offset and amplitude variations; both increase the difference between fringes, which manifests as spurious contributions to the IAW signal. In contrast, the LAMP method gives a comparably low LOD with and without offset or amplitude variations. This reduced sensitivity to noise is a result of the Morlet wavelet convolution, which is an effective technique for filtering thin film interference fringes because it acts as a very strict band pass filter. The shape of a Morlet wavelet closely resembles the shape of Fabry-Pérot fringes in terms of fundamental frequency constituents.

Other than noise immunity, complex Morlet wavelet convolution also provides an integrated way of extracting and therefore normalizing amplitude variations; other filtering methods require an additional step, such as applying the Hilbert transform. Furthermore, the inclusion of an FFT in the signal processing pipeline has another advantage, which is a coarse frequency indicator enabling an unlimited measurement range; the IAW method does not include this and is thus limited to half the free spectral range.

While Table 1 compares only RIFTS, IAW, and LAMP, additional signal processing approaches were considered. Other techniques such as summing of the cross-correlation between a reference and shifted spectra, least squares fitting of the Fabry-Pérot fringe functional form (Equation 1), super-resolution frequency estimation techniques such as the multiple signal classification (MUSIC) algorithm, and fast impulse response (FIR) bandpass filtering in conjunction with the Hilbert transform were investigated, but did not perform as well as the LAMP method.

**Table 1. LOD in refractive index units (RIU) for several signal processing techniques applied to single layer PSi reflectance data generated computationally using the transfer matrix method with added noise.**

| | Simple refractive index change | Refractive index change with linearly varying offset | Refractive index change with linearly varying amplitude |
|---|---|---|---|
| RIFTS | $8.2 \times 10^{-4}$ | $2.5 \times 10^{-3}$ | $8.9 \times 10^{-4}$ |
| IAW | $3.9 \times 10^{-4}$ | $1.7 \times 10^{-3}$ | $1.9 \times 10^{-2}$ |
| LAMP | $5.7 \times 10^{-5}$ | $1.3 \times 10^{-4}$ | $7.4 \times 10^{-5}$ |

While the computational complexity of the LAMP method may initially appear to be a potential concern for future point-of-care applications, given the combined steps of a FFT, wavelet convolution, and phase difference calculation, embedded implementation of real time FFT and wavelet convolution algorithms have been realized[29]. Accordingly, no insurmountable difficulty is anticipated in designing an instrument using these techniques that can update its readout every few seconds, particularly since such computation time is negligible compared to sensor response times.

Further reduction in achievable LOD with the LAMP technique could be achieved by using a thicker film, optimizing wavelet parameters, and tuning the range of filtered data. Thicker films have higher frequency Fabry-Pérot fringes and an increased optical thickness, which enables better discrimination between the desired signal frequency and unwanted low frequency noise contributions. However, thick PSi layers may suffer from mass transport challenges that result in longer response times and sensitivity limitations. Mitigation of these challenges may be possible by using larger pore diameters, mixing approaches, and flow-through configurations[17,18]; however, selecting the most appropriate PSi thickness ultimately depends on a balance between noise mitigation and mass transport limitations. Lower LOD may also be achievable by optimizing the width of the wavelet. In this work, the width of the wavelet was intuitively selected to be equivalent to the FWHM of the dominant FFT peak to enable maximum generalizability to any given system. However, there is no fundamental reason why choosing this width would provide the best rejection of noise while retaining the maximum useful portion of the signal. In this investigation, the range of spectral data filtered and processed is the same for all signal processing techniques. However, in practice some sensing systems will benefit from filtering a wider range of spectral data, but then discarding sections of filtered data at the highest and lowest wavenumber values before processing and obtaining the LAMP result. Finally, the resolution of the interpolated data and wavelet will improve accuracy at the expense of computational complexity, which is a tradeoff governed by sensor performance requirements.

### BSA Assay

The LOD achievable through application of each of the signal processing methods to an experimentally obtained data set was investigated by exposing PSi thin film sensors to a series of BSA concentrations, from 0 to 300 μM. The spectra were analyzed, and several adsorption isotherms were fit to the data using nonlinear regression to determine the most appropriate adsorption model.

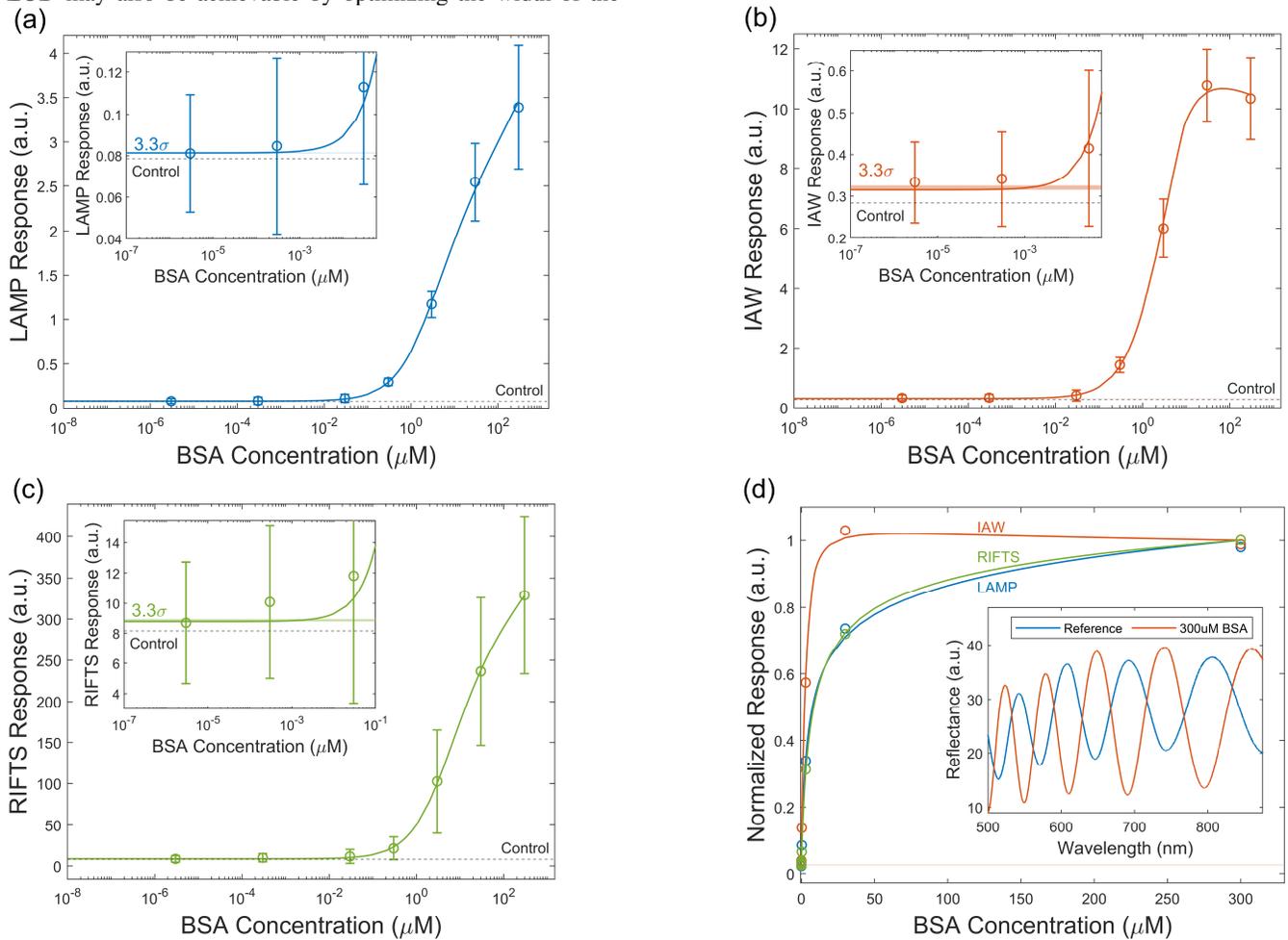

Figure 2. Results of exposing a single layer of oxidized PSi to solutions of BSA in buffer, with concentrations between 0 and 300 μM, analyzed using (a) LAMP, (b) IAW, and (c) RIFTS signal processing methods, shown on a semi-log plot. A Redlich-Peterson adsorption isotherm line of best fit is overlaid, as well as the noise floor for each method. The insets show the data, noise floor and trendline around in the region of the LOD. The normalized responses of all signal processing techniques are shown on a linear scale (d), the inset is an illustration of the maximum shift caused by 300 μM BSA exposure.

**Table 2. LOD for BSA exposure analyzed using IAW, RIFTS and LAMP signal processing methods, showing noise level and Redlich-Peterson adsorption isotherm fit parameters.**

|  | $3.3\sigma$ (a.u.) | LOD (nM) | $I$ (a.u.) | $A$(LM$^{-1}$) | $B$(LM$^{-1}$) | $\beta$ |
|---|---|---|---|---|---|---|
| RIFTS | $1.62 \times 10^{-1}$ | 3.24 | 8.78 | 50.16 | 0.24 | 0.92 |
| IAW | $1.00 \times 10^{-2}$ | 2.50 | 0.32 | 4.02 | 0.37 | 0.95 |
| LAMP | $1.85 \times 10^{-4}$ | 0.22 | 0.08 | 0.82 | 0.47 | 0.88 |

We assumed that the measurements were independent and normally distributed, while recognizing the heteroscedasticity of the data. As a result, the adsorption model fit required the minimization of the sum of the squared residuals weighted by the inverse of the variance, according to maximum likelihood estimation.

The Redlich-Peterson adsorption isotherm, which is applied to calculate the trend lines shown in (Figure 2), was the best fit to the data, as determined by calculating the reduced chi squared goodness of fit metric. This adsorption isotherm model implies imperfect monolayer adsorption. The Redlich-Peterson isotherm is of the form shown in Eqn. (3)

$$\theta = I + \frac{AC_e}{1 + BC_e^\beta} \quad (3)$$

where $\theta$ is the response of the signal due to biomolecule adsorption and the fitting parameters are: $I$ is the intercept or predicted response to a pure buffer control, $A$ is the Redlich-Peterson isotherm constant, $B$ is a constant, $\beta$ is constant between 0 and 1, and $C_e$ is the equilibrium concentration of BSA solution. The IAW line of best fit is complicated by the inherent nonlinearity of the method, the IAW response decreases when the BSA concentration is increased from 3 to 300 μM (Figure 2 (b)), which the LAMP and RIFTS responses show is nearing the saturation region (Figure 2 (a) and (c)). The apparently different trend for IAW can be accounted for by observing that the largest spectral shift for a concentration of 300 μM, exceeds the free spectral range limit of the IAW method (Figure 2 (d) inset). To take the nonlinearity into account, the deviation from a linear response as a function of percentage EOT change was determined by simulation using the transfer matrix method and was fit with a second order polynomial. The adsorption isotherm was then scaled by this nonlinearity using the percentage EOT change provided by the RIFTS result, but was otherwise fit to the data in the same way as the other methods. The LOD was determined by finding the concentration at which the Redlich-Peterson adsorption isotherm line of best fit exceeds its intercept by $3.3\sigma_{blank}$, the noise floor in the blank. The fitting parameters and LOD for the RIFTS, IAW and LAMP methods are shown in Table 2, and the data, the Redlich-Peterson adsorption isotherm line of best fit and the noise floor for each method are shown in Figure 2. The Redlich-Peterson adsorption models simplifies to the Langmuir isotherm when $\beta = 1$. For all three signal processing methods, the optimal value for $\beta$ is close to 1, suggesting only a small perturbation from ideal monolayer formation on a homogenous surface.

The relative differences in LOD values between the signal processing techniques seen experimentally (Table 2) are in good agreement with the theoretical predictions of Table 1. Since the BSA assay is a simple experiment in buffer solution, the presence of noise signatures is minimal and compares favorably with the simple refractive index change case in Table 1. The lowest LOD achievable for BSA is 222 pM (4.44 femtomoles in a 20 μL solution) when the LAMP method is applied, which is one order of magnitude lower than for IAW or RIFTS methods. The LOD values shown in Table 2 are not as low as those recently reported in the literature for a PSi sensor[21]. The predominant reason lies in the method of calculating LOD. Common practice is to center the noise floor around 0, rather than centering the noise floor around the intercept of the line of best fit. However, we believe centering the noise floor around 0 is misleading because if the intercept is increased, the LOD can be artificially decreased and, moreover, the LOD becomes undefined when the intercept is outside the noise floor. Hence, the method we use for calculating LOD in this paper is a more conservative, and importantly more robust, approach.

## Comparison of Different Signal Processing Techniques Applied to Experimental Data

### AGR2 Assay

The robustness of the LAMP method to low frequency offset and amplitude variations, in comparison to the RIFTS and IAW methods, was investigated through the application of the signal processing techniques to experimental data of an aptamer-based single-layer PSi biosensor. This biosensor is designed to selectively target ARG2 protein and consists of oxidized PSi Fabry-Pérot thin films, in which amino-terminated anti-AGR2 aptamers[26] are immobilized within via NHS/EDC coupling chemistry[28]. Reflectance measurements of the biosensor upon exposure to target ARG2 protein and non-target IgG protein (at a concentration of 100 μg mL$^{-1}$ and 200 μg mL$^{-1}$, respectively) in buffer solution and 100 μg mL$^{-1}$ AGR2 in plasma were carried out. The reflectance data shown in the insets of Figure 3 reveal that both offset and amplitude variations are present in the measured data. We note that the magnitude of the RIFTS, IAW, and LAMP processed signals were normalized to have the same minimum and maximum values for ease of comparison. Common trends are shown in the data processed with RIFTS, IAW, and LAMP: (1) a relatively stable baseline value is established during buffer solution incubation, (2) a large increase in signal magnitude occurs when the ARG2, IgG protein, or plasma alone is introduced due to the change in the refractive index of the introduced solution compared to the baseline buffer, (3) a more gradual or no additional signal change occurs as the protein is incubated on top of the PSi and diffuses into the porous layer, while binding to the immobilized aptamers on the pore walls, (4) a signal reduction occurs when a buffer rinse is carried out to remove unbound protein from the pores, and (5) a relatively stable new baseline value is established in most cases upon incubating the sample with buffer solution once again.

Key differences in S/N, ability to resolve protein capture dynamics, and accuracy of the biosensing result can be seen by comparing the signals processed by RIFTS, IAW, and LAMP methods in Figure 3. When using RIFTS, the noisy signal makes it difficult to determine the magnitude of the signal change due to protein capture with high accuracy and also makes it more challenging to ascertain target protein capture dynamics. Low frequency noise signatures lead to false trends for protein capture in IAW data, which is of significant concern for the accuracy of the PSi biosensor. In Figure 3 (b), the IAW method suggests that there is a large signal change when the sensor is exposed to the non-target protein, giving a false positive result. In Figure 3 (d), artefacts in the IAW signal are present, due to the IAW response to predominantly offset changes. Moreover, in Figure 3 (a,c), the IAW method suggests an unexpected continually increasing signal during the post-protein exposure buffer rinse and incubation, which is not consistent with the RIFTS and LAMP results and is physically misleading; while positive increase in the optical signal is attributed to molecules diffusion into the porous layer and the corresponding increase in its average refractive index, no protein is present in the solution on top of the biosensor at this stage[18]. It is important to note that because the IAW processed response can be dominated by the change in offset and amplitude of the reflectance spectra, instead of by frequency shifts that are present due to molecular attachment, caution must be taken to eliminate systematic noise sources in experiments when utilizing the IAW method. Otherwise, it is possible that the IAW result may be disconnected from the physics of the system, as exemplified in Figure 3 (c,d).

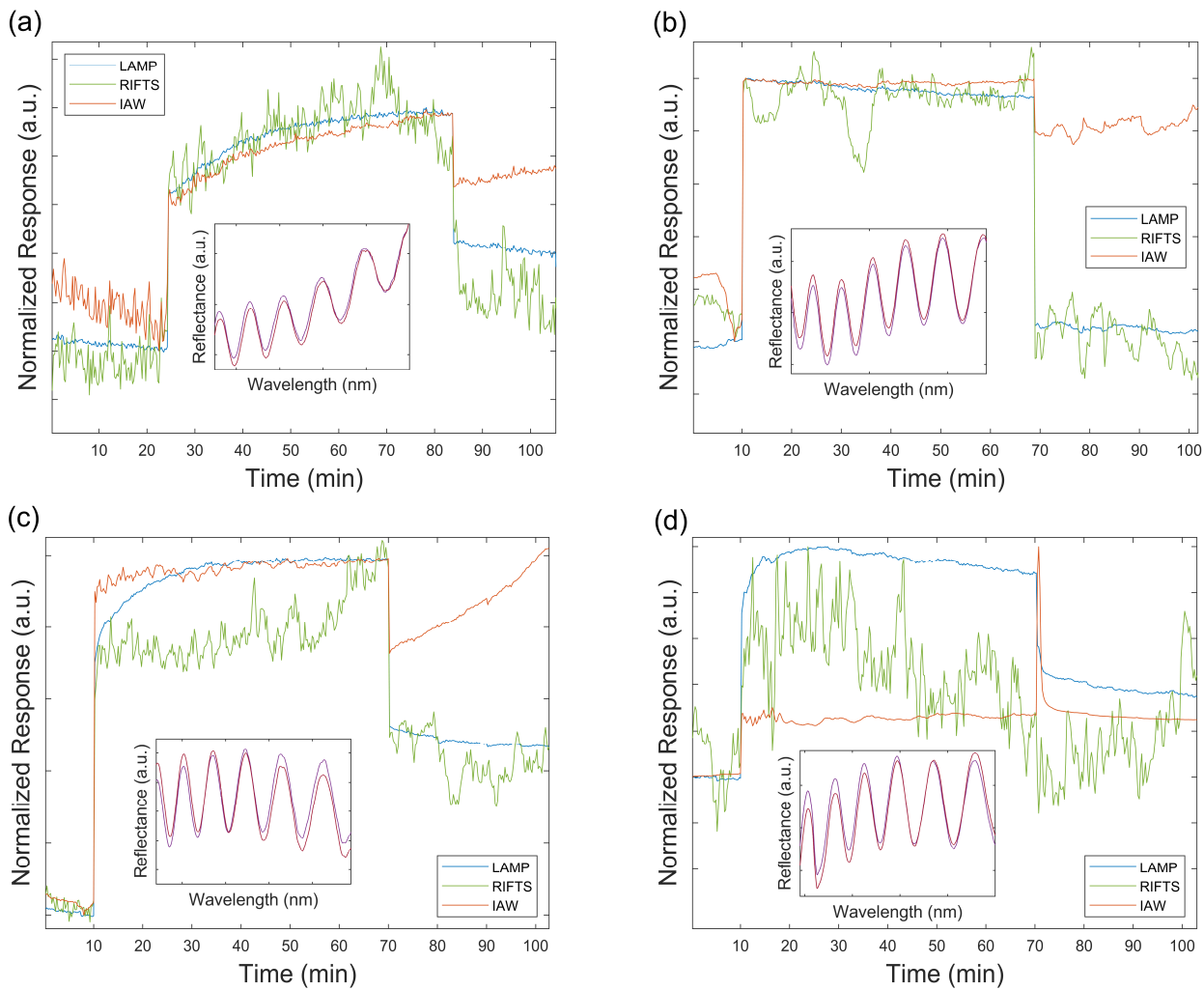

Figure 3. Comparison of signal processing techniques applied to experimental data collected by exposing the single-layer PSi biosensor to (a) 100 µg mL$^{-1}$ ARG2, the target protein, in buffer solution, (b) 200 µg mL$^{-1}$ non-target protein IgG in buffer, (c) 100 µg mL$^{-1}$ ARG2 spiked in 50% plasma in buffer, and (d) 50% neat plasma in buffer. In all cases, there is a pre- and post-wash in buffer. Insets show reflectance spectra taken during the pre- and post-wash steps; signal offset and amplitude variations (i.e., noise signatures) can be observed.

In all cases studied in Figure 3, the LAMP method is shown to be the most robust and reliable signal processing approach. There is a clear distinction between signal changes for target protein capture, negligible signal change when the biosensor is exposed to a non-target protein in buffer, and a small non-specific binding induced signal change when the biosensor is exposed to non-spiked plasma. While the LAMP and RIFTS signal processing techniques lead to similar final results for most biosensing experiments, the LAMP signal has a much higher S/N ratio than the RIFTs signal, which should lead to a lower experimentally demonstrated LOD. Moreover, the LAMP method clearly elucidates the target protein capture dynamics.

## Conclusion

Applying the LAMP technique, which is based on complex Morlet wavelet band pass filtering and calculation of average phase difference, to PSi biosensor reflectance data provides a robust and sensitive measure of frequency shifts that arise due to biomolecular recognition and binding events in the pores. The key benefits of the LAMP method are reducing the effect of noise present across the frequency spectrum and mitigating spurious signal changes that otherwise result from low frequency variations in the offset and amplitude of measured spectra. Analysis of both simulated data and experimental data from a BSA assay demonstrate that the LAMP method achieves a LOD that is approximately one order of magnitude lower than that of RIFTS and IAW – two other signal processing approaches used to analyze data from PSi sensors and biosensors. Moreover, it was shown that the LAMP method can be reliably applied to noisy data acquired in complex biological media. Improved S/N and selectivity were demonstrated by applying LAMP to data from an aptamer-based biosensor for targeting a cancer biomarker, AGR2 protein, compared to the application of RIFTS and IAW to the same data set. The robustness and lower detection limit achievable by employing the LAMP signal processing technique can benefit many applications where noise in data sets cannot be perfectly controlled and low detection limits are required for clinical relevancy.

# AUTHOR INFORMATION

## Corresponding Author

dummydummydummydummydummydummy* Email: simon.j.ward@vanderbilt.edu.

## Author Contributions

All authors have given approval to the final version of the manuscript.

## Funding Sources

NIH grant # R21AI156693 and the ISRAEL SCIENCE FOUNDATION (grant No. 704/17).

## Notes

The authors declare no competing financial interest.

# ACKNOWLEDGMENT

This work was supported in part by the NIH (R21AI156693). The authors thank Dr. Catie Chang and Dr. D. Mitchell Wilkes for helpful discussions related to signal processing.

# ABBREVIATIONS

LOD, PSi, RIFTS, FFT, IAW, LAMP, FWHM, S/N, AGR2, HF, PDMS, IgG, SB, SEM, DI, BSA, MUSIC, FIR.